%% file: eprint.tex
\documentclass[12pt]{article}
\usepackage{graphicx}
\usepackage{amsmath, amsfonts, xfrac}
\usepackage[displaymath, mathlines]{lineno}

\newcommand\pubnumber{NuPhys2015-Martins}
\newcommand\pubdate{\today}

\def\napoli{School of Physics and Astronomy\\
Queen Mary University of London, E1 4NS London, UK}

\def\Title#1{\begin{center} {\Large #1 } \end{center}}
\def\Author#1{\begin{center}{ \sc #1} \end{center}}
\def\Address#1{\begin{center}{ \it #1} \end{center}}

\newcommand\pubblock{\rightline{\begin{tabular}{l} \pubnumber\\
         \pubdate  \end{tabular}}}
\newenvironment{Abstract}{\begin{quotation}  }{\end{quotation}}
\newenvironment{Presented}{\begin{quotation} \begin{center} 
             PRESENTED AT\end{center}\bigskip 
      \begin{center}\begin{large}}{\end{large}\end{center} \end{quotation}}

\input econfmacros.tex

\begin{document}
\begin{titlepage}
\pubblock

\vfill
\Title{Charged Current Coherent Pion Production in Neutrino Scattering}
\vfill
\Author{Paul Martins, for the T2K Collaboration} 
\Address{\napoli}
\vfill
\begin{Abstract}
We summarise here the main differences of three models of neutrino-induced coherent pion production, namely the Rein-Sehgal and Berger-Sehgal models based on the Partially Conserved Axial Current theorem and the Alvarez-Ruso \textit{et al.} model which is using a microscopic approach. Their predictions in the event generators are compared against recent experimental measurements for a neutrino energy from 0.5 to 20\,GeV.
\end{Abstract}
\vfill
\begin{Presented}
NuPhys2015, Prospects in Neutrino Physics\\
Barbican Centre, London, UK,  December 16--18, 2015
\end{Presented}
\vfill
\end{titlepage}
\def\thefootnote{\fnsymbol{footnote}}
\setcounter{footnote}{0}
%
\section{Introduction}  
With the development of accelerator-based long-baseline neutrino oscillation experiments, it quickly became essential to have a better understanding of neutrino-nucleus scattering at the neutrino energies of $\sim$1\,GeV in order to reduce the systematic uncertainties on oscillation parameter measurements. Hence new cross-section measurements for various interaction modes and different nuclear targets have been carried out in the last decade. One important channel to consider is coherent pion production. It occurs when a neutrino scatters off a nucleus and leaves the target in its ground state, only producing one lepton and one pion. Today, neutral current (NC) coherent $\pi^{0}$ production remains a background source for the $\nu_e$ appearance oscillation channel. Further studies were also motivated by the K2K and SciBooNE searches for charged current (CC) coherent $\pi^{+}$ production that had no evidence of such a process for neutrino energy below 2\,GeV, while it was already observed for higher energies.

\section{Theoretical models}
\subsection{PCAC based models}

The Adler's PCAC theorem states that, for a weak interaction process where the outgoing lepton is parallel to the incoming neutrino (small angle approximation), the scattering amplitude only relies on the divergence of the axial-vector current \cite{Adler:1968tw}, which can be estimated through the pion decay scattering amplitude. Therefore the neutral-current neutrino coherent pion production differential cross-section can be related to the differential pion-nucleus cross-section $d\sigma / d\vert t \vert$:
\begin{center}
$
\dfrac{d\sigma}{dx\,dy\,d\vert t \vert} = \dfrac{G^{2}\,M\,E_{\nu}}{2\,\pi^{2}} f_{\pi}^{2}\,(1-y) \times \dfrac{d\sigma (\pi^{0} \mathfrak{N}\to \pi^{0} \mathfrak{N})}{d\vert t \vert}
$
\end{center}
where $x=Q^{2}/2MyE_{\nu}$ and $y=(E_\nu - E_{lep})/E_{\nu}$ are the Bjorken scaling variables, $M$ is the nucleon mass and $f_{\pi}$ is the pion decay constant. Although this equation should only be valid for $Q^{2}=0$, a dipole propagator was used in \cite{Rein:1982pf,Berger:2008xs} to extend it to non-forward directions. The Rein-Sehgal model \cite{Rein:1982pf} further relates the pion-nucleus cross-section in terms of the total and inelastic pion-nucleon cross-sections via the optical theorem:
\begin{center}
$
\dfrac{d\sigma (\pi^{0} \mathfrak{N}\to \pi^{0} \mathfrak{N})}{d\vert t \vert} = A^{2}\dfrac{1}{16\pi}\left[\sigma_{tot}^{\pi^{0}N}\right]^2 (1+r^2)\, \text{exp}\left(\dfrac{-R_{0}^2 A^{2/3} \vert t \vert}{3} \right) \text{exp}\left(\dfrac{-9A^{1/3} \sigma_{inel}}{16 \pi R_{0}^{2}}\right)
$
\end{center}
with $A$ the target atomic number, $r$ the forward scattering amplitude ratio and $R_0$ the nucleon radius. The Berger-Sehgal model \cite{Berger:2008xs} improvements include using the available data on differential and total pion-carbon cross-sections, which makes it a better approximation for neutrino-carbon scattering. Aditionally, the pion-carbon scattering cross-section is greatly reduced compared to pion-nucleus RS approximations in the E$_\pi< $ 1\,GeV region. Thus the model improves predictions for pion energies below 1\,GeV. In case of CC interactions, both models take into account a phase space correction due to the non-negligible lepton mass \cite{Rein:2006di}.

\subsection{Microscopic aproach}
The Alvarez-Ruso \textit{et al.} model \cite{AlvarezRuso:2007it} is an example of a microscopic model (i.e. at the neutrino-nucleon interaction level). It considers the neutrino-nucleon pion production processes where the nucleon remains in its ground state. All the amplitudes are then summed coherently, but the s-channel $\Delta$ mode is largely dominant. This model takes into account nuclear effects that modify the $\Delta$ properties (Pauli-blocking, absorption) which, given the model's assumptions, reduces its validity to neutrino energies below 3\,GeV.

\section{Comparison against recent measurements}
\begin{figure}[h]
\begin{center}
\includegraphics[height=0.3\linewidth]{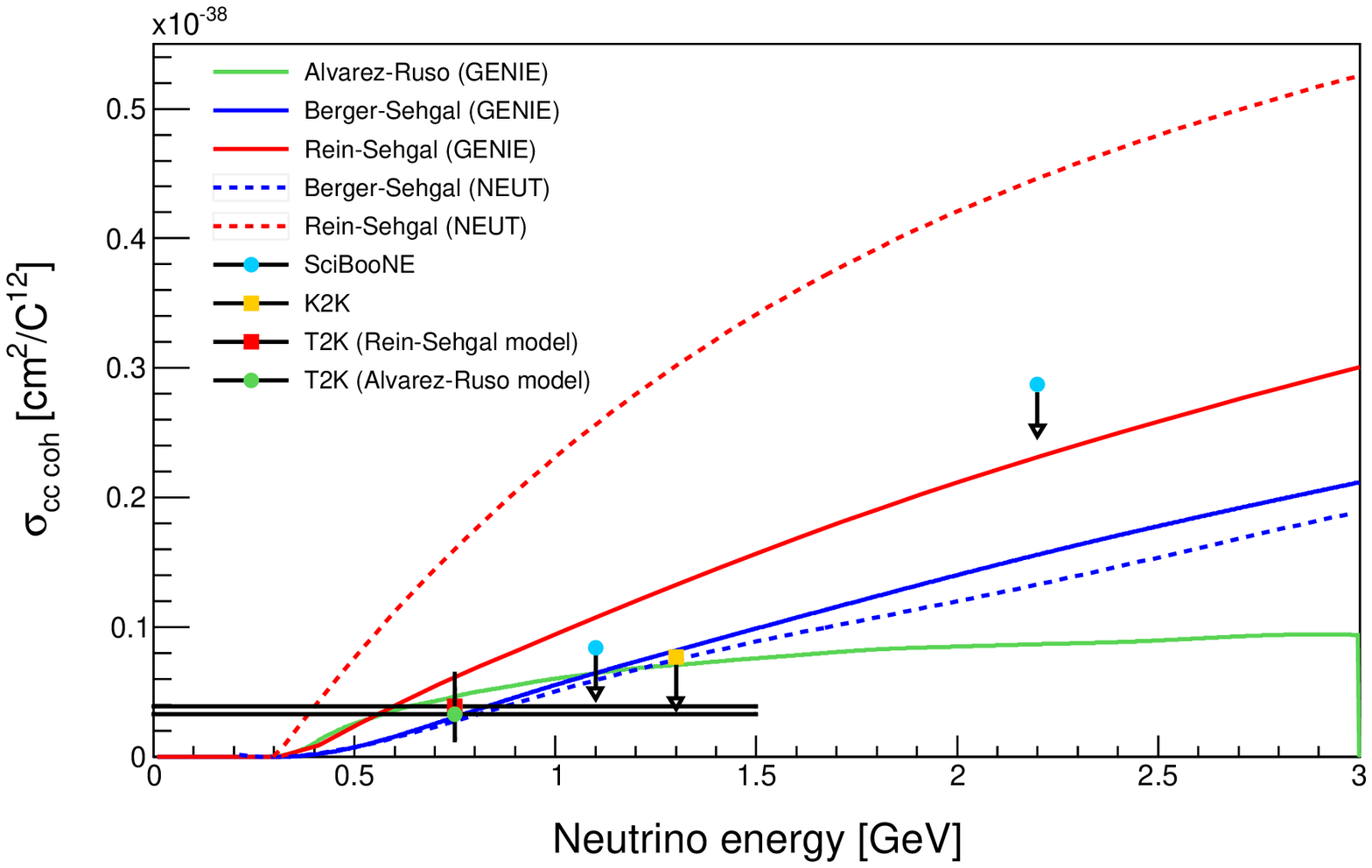}
\includegraphics[height=0.3\linewidth]{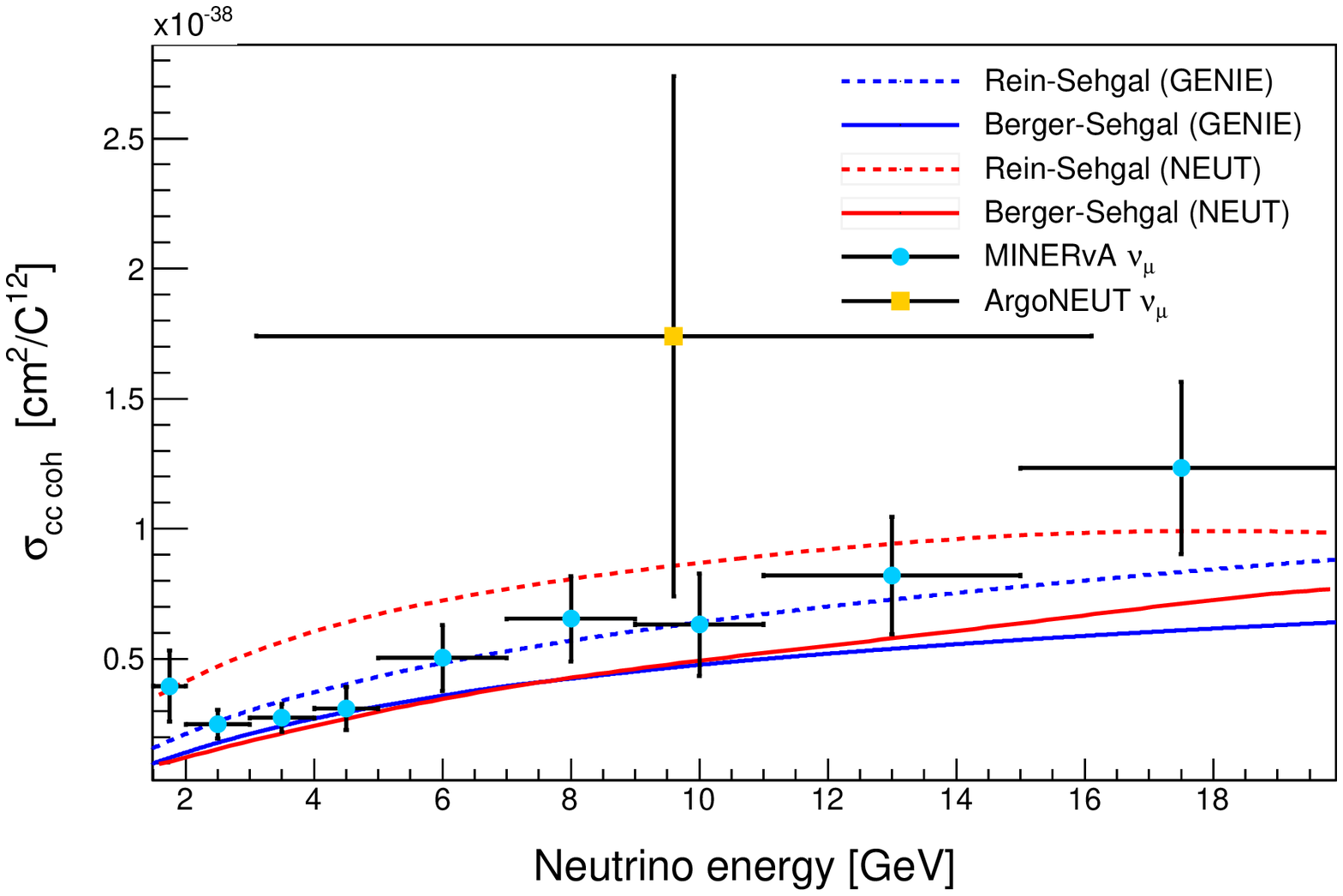}
\end{center}
\caption{Left: comparison of the Rein-Sehgal (red), Berger-Sehgal (blue) and Alvarez-Ruso \textit{et al.} (green) models as predicted by NEUT (dashed) and GENIE (solid) against the K2K, SciBooNE and T2K results. Right: comparison of the Rein-Sehgal and Berger-Sehgal models against the MINER$\nu$A and ArgoNEUT results}
\label{fig:lowE}
\end{figure}

A few experiements had already measured the $\nu_\mu$ CC coherent cross-section for nuclear targets above A = 20 (Neon) and neutrino energies above 7\,GeV \cite{Grabosch:1985mt,Allport:1988cq,Vilain:1993sf,Aderholz:1988cs}. In 2005 and 2009, K2K and SciBooNE respectively conducted the first searches for CC coherent interactions using a carbon target and an average neutrino energy around 1\,GeV. Their results didn't indicate conclusive evidence of such a process in this energy range and limits on the cross-section were set \cite{Hasegawa:2005td,Hiraide:2009zz}. It is worth recalling that these two experiments used the Monte-Carlo predictions using the Rein-Sehgal model as implemented in NEUT \cite{Hayato:2009zz}, which predicts roughly twice the interaction rate in the range 0\,GeV $<$ $E_\pi$ $<$ 1.0\,GeV than the Berger-Sehgal model. This effect is shown in Fig.\ref{fig:lowE}, where the GENIE \cite{Andreopoulos:2009rq,Andreopoulos:2015wxa} predictions are also given.

For low neutrino energies, SciBooNE and K2K upper limits and the recent T2K measurement \cite{Abe:2016fic} are compatible with the Berger-Sehgal and Alvarez-Ruso \textit{et al.} models . A similar comparison is shown for neutrino energies up to 20\,GeV with the MINER$\nu$A \cite{Higuera:2014azj} and ArgoNEUT \cite{Acciarri:2014eit} results, where the latter was scaled to a carbon target using a $A^{1/3}$ rule\footnote{this is motivated by the global $A^{1/3}$ dependance of the Rein-Sehal cross-section}.

We notice a large discrepancy between GENIE and NEUT for the Rein-Sehgal model that could be explained by the different implementations in the generators in terms of the phase space coverage in the $\vert t \vert$ integral, the scattering amplitude ratio $r$ and the pion-nucleon scattering data.

\begin{figure}[h]
\begin{center}
\includegraphics[height=0.3\linewidth]{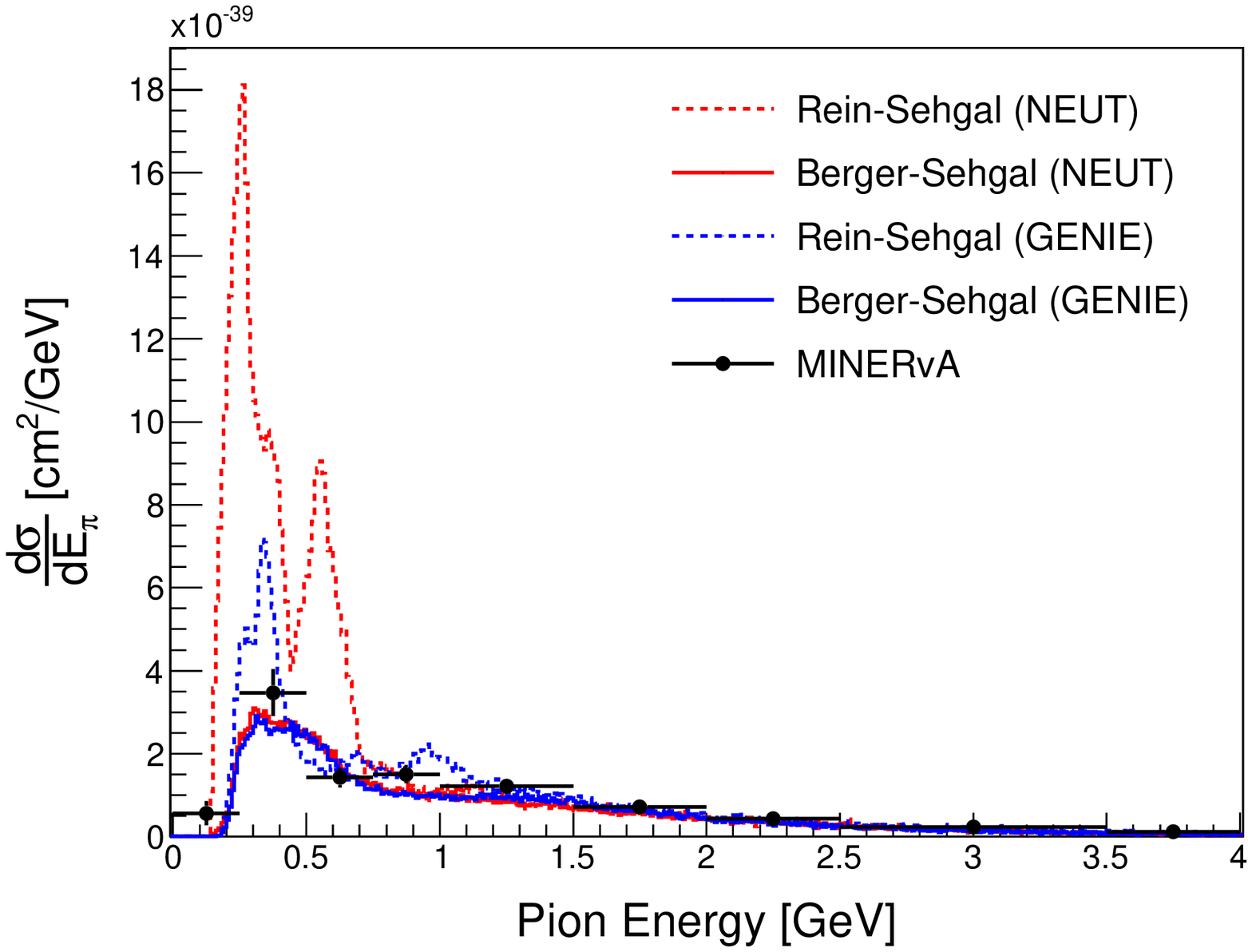}
\includegraphics[height=0.3\linewidth]{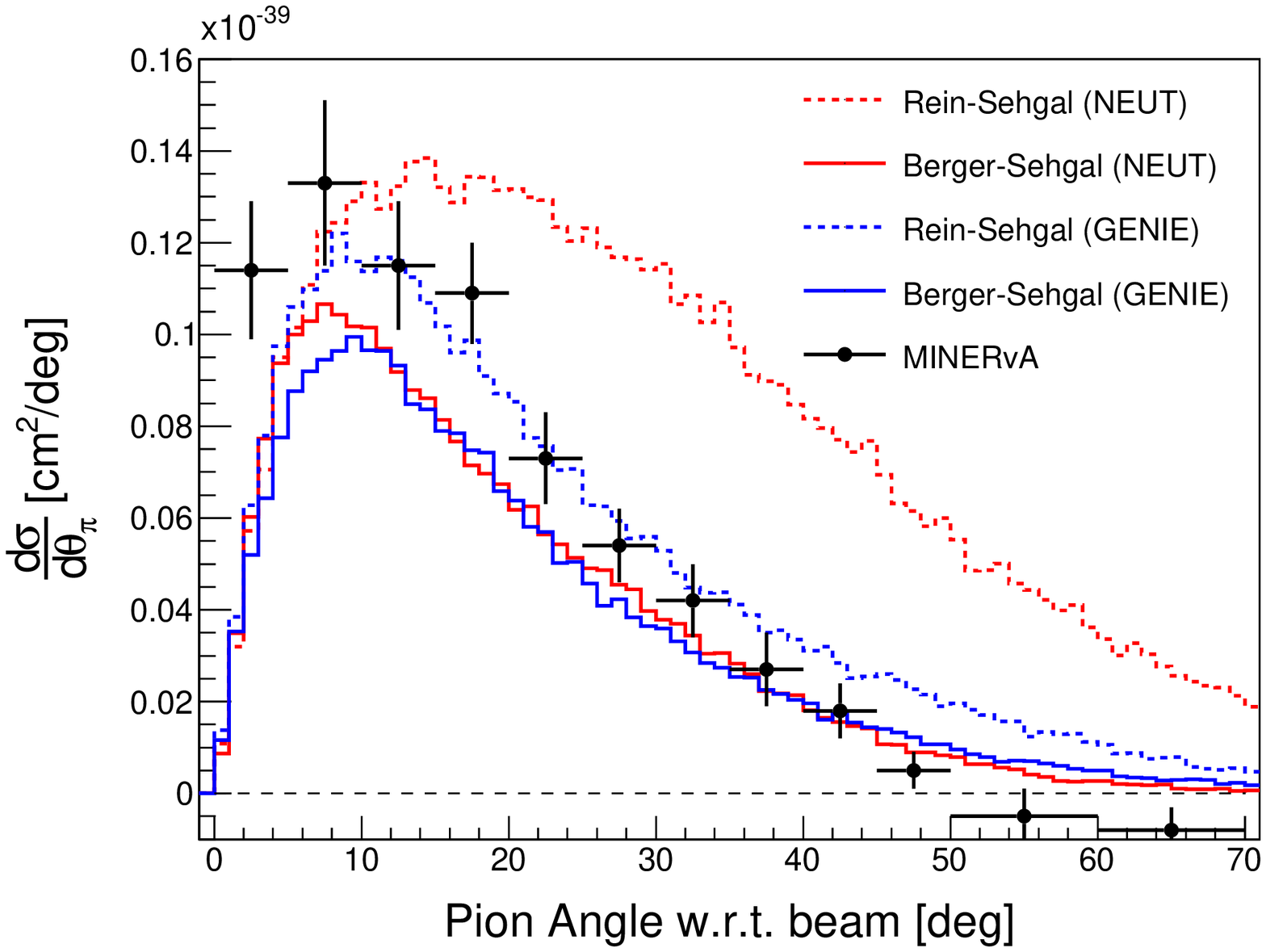}
\end{center}
\caption{Differential cross-section measurements as a function of momentum (left) and angle (right). The MINER$\nu$A results are compared against the Rein-Sehgal (dashed), Berger-Sehgal (solid) predictions in GENIE (blue) and NEUT (red).}
\label{fig:difXS}
\end{figure}

For neutrino energies above 2\,GeV, the MINER$\nu$A experiment measured for the first time the neutrino and anti-neutrino differential CC coherent cross-sections as a function of the pion momentum and pion angle \cite{Higuera:2014azj}. As the mean energy is beyond the validity range of the Alvarez-Ruso \textit{et al.} model, the comparison in Fig.\ref{fig:difXS} only shows the two PCAC based models. We observe a better agreeement of the MINER$\nu$A data with the Berger-Sehgal model, especially in the low pion momentum region. Although both generators have different predictions for the Rein-Sehgal model, they tend to have similar behavior for the Berger-Sehgal model up to a neutrino energy of 13\,GeV, due to the use of the same $\pi$-C scattering data.

\section{Conclusion}
The difference in the pion-nucleus cross-section calculation between the Rein-Sehgal and Berger-Sehgal models have an important influence on their predictions. From the comparison with recent data, it seems clear that the Rein-Sehgal model overestimates the total and differential coherent cross-section, in both NEUT and GENIE generators, while the agreement with Berger-Sehgal is better. The limits on the total cross-section set by the K2K and SciBooNE searches are compatible with both Berger-Sehgal and Alvarez-Ruso \textit{et al.} models. Finally, while the latest T2K measurement cannot distinguish between microscopic and PCAC based models, it does remove any doubt as to whether CC neutrino coherent interactions occur for neutrino energies below 3\,GeV.


\end{document}

%% file: econfmacros.tex
\def\beq{\begin{equation}}
\def\eeq#1{\label{#1}\end{equation}}
\def\eeqn{\end{equation}}

\def\beqa{\begin{eqnarray}}
\def\eeqa#1{\label{#1}\end{eqnarray}}
\def\eeqan{\end{eqnarray}}

\let\bar=\overbar

\def\Dslash{\not{\hbox{\kern-4pt $D$}}}
\def\dslash{\not{\hbox{\kern-2pt $\del$}}}

\def\msb{{\bar{\ssstyle M \kern -1pt S}}}

